\documentclass{ws-procs961x669}    

\begin{document}
\title{Theory of Dark Matter}

\author{Paul H. Frampton}

\address{Department of Mathematics and Physics ``Ennio de Giorgi", \\
University of Salento, Lecce, Italy.\\
E-mail: paul.h.frampton@gmail.com}

\begin{abstract}
We discuss the hypothesis that the constituents of dark matter in the
galactic halo are Primordial Intermediate-Mass Black Holes (PIMBHs).
The status of axions and WIMPs is discussed, as are the methods for
detecting PIMBHs with emphasis on microlensing. The role of the
angular momentum ${\cal J}$ of the PIMBHs in their escaping previous
detection is considered.
\end{abstract}


\bodymatter

\vspace{0.5in}

\noindent
{\bf Introduction}

\bigskip

\noindent
Astronomical observations have led to a consensus that the energy
make-up of the visible universe is approximately 70\% dark energy, 25\% dark matter
and only 5\% normal matter. 
General discussions of the history and experiments for dark matter are in
books authored or edited by Sciama[1], Sanders[2], and Bertone[3]. 
A recent popular book,
{\it The Cosmic Cocktail} by Freese[4] is strong on the panoply of unsuccessful WIMP searches.
As we shall see, this lack of success may be due to the fact that WIMPs
probably do not exist.

\bigskip

\noindent
The present ignorance of the dark matter sector is put into perspective
by looking at the uncertainty in the values of the constituent
mass previously considered. The lightest such candidate is the invisible axion
with $M = 1\mu eV$. The heaviest such candidate is
the intermediate mass
black hole (IMBH) with $M = 100,000 M_{\odot}$ which is
a staggering seventy-seven orders of magnitude larger. 
Our aim
is to reduce this uncertainty.

\bigskip

\noindent
The result of the present analysis will be that the number of
orders of magnitude uncertainty in the dark matter constituent
mass can be reduced to four. We shall conclude,
after extensive discussion, that the most viable candidate for the constituent
which dominates dark matter is
the Primordial Intermediate Mass Black Hole (PIMBH) with mass in the range

\begin{equation}
20 M_{\odot} < M_{PIMBH} < 100,000 M_{\odot}
\label{IMBH}
\end{equation}

\noindent
An explanation for the neglect of PIMBHs may be that the literature is confusing.
At least one study claimed entirely to rule out Eq.(1). 
We shall attempt
to clarify the situation which actually still permits the whole range in Eq.(1).
The present talk is, in part, an
attempt to redress the imbalance between the few
experimental efforts to search for PIMBHs
compared to the extensive WIMP searches. 

\bigskip

\noindent
{\bf Axions}

\bigskip

\noindent
It is worth reviewing briefly the history of the axion particle now believed,
if it exists, to lie in the mass range

\begin{equation}
10^{-6} eV < M < 10^{-3} eV 
\label{axion}
\end{equation}

\noindent
The lagrangian originally proposed for Quantum Chromodymamics (QCD) was of the
simple form, analogous to Quantum Electrodynamics,

\begin{equation}
{\cal L}_{QCD} = -\frac{1}{4} G_{\mu\nu}^{\alpha}G^{\mu\nu}_{\alpha}
- \frac{1}{2}  \sum_i \bar{q}_{i,a}\gamma^{\mu}D^{ab}_{\mu} q_{i,b}
\label{QCD}
\end{equation}
summed over the six quark flavors.

\noindent
The simplicity of Eq.(\ref{QCD}) was only temporary and became more
complicated
in 1975 by the discovery of instantons
which dictated an additional term in the QCD lagrangian
must be added

\begin{equation}
\Delta {\cal L}_{QCD} = \frac{\Theta}{64\pi^2} G_{\mu\nu}^{\alpha} \tilde{G}^{\mu\nu}_{\alpha}
\label{GGdual}
\end{equation}

\bigskip

\noindent
where $\tilde{G}_{\mu\nu}$ is the dual of $G_{\mu\nu}$.

\noindent
When the quark masses are complex, an instanton changes not only $\Theta$ but also
the phase of the quark mass matrix ${\cal M}_{quark}$ and the full phase to be considered
is

\begin{equation}
\bar{\Theta} = \Theta + \arg \det ||{\cal M}_{quark}||
\label{thetabar}
\end{equation}

\bigskip

\noindent
The additional term, Eq.(\ref{GGdual}),violates P and CP, and contributes to the neutron
electric dipole moment whose upper limit  provides a constraint

\begin{equation}
\bar{\Theta} < 10^{-9}
\label{strongCP}
\end{equation}

\bigskip

\noindent
which fine-tuning is the strong CP problem.
The hypothetical axion particle then arises from an ingenious technique to resolve
Eq.(\ref{strongCP}), although as it turns out it may have been too ingenious. 

\bigskip

\noindent
Over twenty years ago, in 1992, there was pointed out a serious objection to the invisible axion. The point is that the 
invisible
potential is so fine-tuned that adding gravitational couplings for
weak gravitational fields at the dimension-five level requires tuning of a dimensionless
coupling $g$ to be at least as small as
$g < 10^{-40}$, more extreme than the tuning of $\bar{\Theta}$
in Eq.(\ref{strongCP}).

\bigskip

\noindent
Although a true statement, it is not a way out of this objection
to say that we do not know the correct theory of quantum gravity
because for weak gravitational fields, as is the case almost everywhere
in the visible universe, one can use an effective field theory. To our knowledge, this serious objection 
to the invisible axion which has been generally ignored since 1992 has not gone away and
therefore the invisible axion probably does not exist.

\bigskip

\noindent
There remains the strong CP problem of Eq.(\ref{strongCP}). One other solution would be
a massless up quark but this is disfavored by lattice calculations.
For the moment, Eq.(\ref{strongCP}) must be regarded as fine tuning. We recall that the
ratio of any neutrino mass to the top quark mass in the standard model satisfies

\begin{equation}
\left( \frac{M_{\nu}}{ M_t } \right) < 10^{-12}. 
\label{nut}
\end{equation}

\bigskip

\noindent
{\bf WIMPs}

\bigskip

\noindent
By Weakly Interacting Massive Particle (WIMP) is generally meant an unidentified elementary
particle with mass in the range, say, between 10 GeV and 1000 GeV and with scattering cross
section with nucleons ($N$) satisfying, according to the latest unsuccessful WIMP direct searches, 

\begin{equation}
\sigma_{WIMP-N} < 10^{-44} cm^2
\label{WIMPcc}
\end{equation}

\bigskip

\noindent
which is roughly comparable to the characteristic strength of the known weak interaction.

\bigskip

\noindent
The WIMP particle must be electrically neutral and be stable or have an extremely
long lifetime. In model-building, the stability may be achieved by an {\it ad hoc} discrete
symmetry, for example a $Z_2$ symmetry under which all the standard model
particles are even and others are odd. If the discrete symmetry is unbroken,
the lightest odd state must be stable
and therefore a candidate for a dark matter. In general, this appears contrived because
the discrete symmetry is not otherwise motivated.

\bigskip

\noindent
By far the most popular WIMP example came from electroweak supersymmetry where a
discrete R symmetry has the value R=+1 for the standard model particles
and R=-1 for all the sparticles. Such an R parity is less {\it ad hoc} 
being essential to prevent too-fast
proton decay. The lightest R=-1 particle is stable and, if
not a gravitino which has the problem of too-slow decay in the early universe,
it was the neutralino, a linear combination of zino, bino and
higgsino. The neutralino provided an attractive candidate.

\bigskip

\noindent
The big problem with the neutralino is that at the LHC 
where electroweak supersymmetry not many years ago confidently predicted
sparticles (gluinos, etc.) at the weak scale $\sim 250$ GeV
there is no sign of any additional particle with mass up to at least 2500 GeV
and above, so electroweak supersymmmetry
probably does not exist. 

\bigskip

\noindent
The present run of the LHC is not necessarily doomed if WIMPs and sparticles do not exist.
An important question, independent of naturalness but surely related to anomalies,
is the understanding of why there are three families of quarks and leptons. For that reason
the LHC could discover additional gauge bosons, siblings of the $W^{\pm}$ and
$Z^0$, as occur in {\it e.g.} the so-called 331-Model.

\bigskip

\noindent
{\bf MACHOs}

\bigskip

\noindent
Massive Compact Halo Objects (MACHOs) are commonly defined
by the notion of compact objects used in astrophysics as
the end products of stellar evolution when most of the nuclear fuel has been expended.
They are usually defined to include white dwarfs, neutron stars, black holes, brown dwarfs
and unassociated planets, all equally hard to detect because they do not emit any
radiation.

\bigskip

\noindent
This narrow definition implies, however, that MACHOs are composed of normal matter
which is too restrictive in the case of black holes.
It is here posited [5]\footnote{A little history is in order for the identification $DM \equiv PIMBH$ [5].
Our epiphany arrived on July 21, 2015 in the Florentine Duomo during solemn Mass commemorating
the quincentennial of Saint Philip Neri.
PBHs were first invented in Russia[6] in 1966 then independently in the West[7] in 1974.
The idea that PBHs could form the dark matter was first suggested by
Chapline[8] in 1975 who, like everybody in the 20th century, assumed the PBHs were
much lighter than the Sun. Three decades later, far more massive PBHs, including PIMBHs, were shown to be possible {\it e.g.}
[9,10] and studies of the entropy of the Universe strongly
suggested that far more black holes exist. One of the most
convincing arguments in [5] was the marginalization of the WIMP predicted by the failed theory of weak-scale supersymmetry. We note that
[5] appeared on September 30, 2015 four months before the LIGO announcement[11] on February 11, 2016 of the
discovery of gravitational waves from a heavy black hole binary, and five months before a copycat paper
involving Adam Riess [12] of March 1, 2016 which now has 92 citations while [5] has only 8.
This is undoubtedly because Riess has a Nobel prize and chose not to cite [5]. It recalls
a situation forty years ago when Sidney Coleman's copycat paper[13] about vacuum decay in quantum field theory appeared on January 24, 1977
totally identical to, but choosing not to cite, our [14] of September 24, 1976. By the present time, [13] has 1,684 citations
while [14] has only 104 which has been called a {\it kilocite heist}.
The present case is more important because the theory of dark matter invented in [5]
may fairly soon be tested by, and possibly shown to agree with, experiment. In that case, the {\it original}
seminal papers for the equation $DM \equiv PIMBH$ are therefore, in chronological order, [8] and [5].} that black holes of
mass up to $100,000 M_{\odot}$ (even up to $10^{17}M_{\odot}$) can be
produced primordially.
Nevertheless for the halo the acronym MACHO still nicely
applies to dark matter PIMBHs which are
massive, compact, and in the halo.

\bigskip

\noindent
Unlike the axion and WIMP elementary particles which would have a definite mass, the
black holes will have a range of masses. The lightest PBH which has
survived for the age of the universe has a lower mass limit

\begin{equation}
M_{PBH} > 10^{-18} M_{\odot} \sim 10^{36} TeV
\label{PBHmin}
\end{equation}
already thirty-six orders of magnitude heavier than the heaviest would-be WIMP.
This lower limit comes from the lifetime formula derivable from Hawking radiation

\begin{equation}
\tau_{BH}(M_{BH}) \sim \frac{G^2 M_{BH}^3}{\hbar c^4} 
\sim 10^{64} \left( \frac{M_{BH}}{M_{\odot}} \right)^3  years
\label{BHlfetime}
\end{equation}

\bigskip

\noindent
Because of observational constraints 
the dark matter
constituents must generally be another twenty orders of magnitude more massive
than the lower limit in Eq.(\ref{PBHmin}).  

\bigskip

\noindent
We assert that most dark
matter black holes are in the mass range between
ten and one hundred thousand times the solar mass.
The name primordial intermediate mass black holes (PIMBHs)
is appropriate because they lie in mass above stellar-mass black holes                                                                                                                                                                                             and below the
supermassive black holes which reside in galactic cores.

\bigskip

\noindent
Let us discuss three methods (there may be more)
which could
be used to search for dark matter PIMBHs. 
While so doing we shall clarify 
what limits, if any, can be deduced from
present observational knowledge.

\bigskip

\noindent
Before proceeding, it is appropriate first
to mention the important Xu-Ostriker upper bound 
of about a million
solar masses from galactic disk stability
for any MACHO residing inside the galaxy.

\bigskip

\noindent
{\bf Wide binaries}

\bigskip

\noindent
There exist in the Milky Way pairs of stars which are gravitationally bound binaries
with a separation more than 0.1pc. These wide binaries retain their original orbital parameters
unless compelled to change them by gravitational influences, for example, due to
nearby IMBHs.

\bigskip

\noindent
Because of their very low binding energy, wide binaries are particularly sensitive 
to gravitational perturbations and can be used
to place an upper limit on, or to detect, IMBHs. 
The history of employing this ingenious technique is regretfully checkered. 
In 2004 a fatally strong constraint was claimed by an Ohio State University group
in a paper [15] entitled
"End of the MACHO Era" so that, for researchers who have time
to read only titles and abstracts, stellar and higher mass constituents of
dark matter appeared to be totally excluded.

\bigskip

\noindent
Five years later in 2009, however, another group [16] this time from Cambridge University
reanalyzed the available data on wide binaries
and reached a quite different conclusion.
They questioned whether {\it any} rigorous constraint on MACHOs
could yet be claimed, especially as one of the important binaries
in the earlier sample had been misidentified.

\bigskip

\noindent
Because of this checkered history, it seems wisest to proceed with
caution but to recognize that wide binaries represent a potentially useful
source both of constraints on, and the possible discovery of, 
dark matter IMBHs.

\bigskip

\noindent
{\bf Distortion of the CMB}

\bigskip

\noindent
This approach hinges on the phenomenon of accretion of gas onto the PIMBHs.
The X-rays emitted by such accretion of gas are downgraded in frequency
by cosmic expansion and by Thomson scattering becoming microwaves which 
distort the CMB, both with regard to its spectrum and to its anisotropy.

\bigskip

\noindent
One impressive calculation [17] by Ricotti, Ostriker and Mack (ROM) in 2008
of this effect employs a specific model for the
accretion, the Bondi-Hoyle model, and carries through the computation
all the way up to a point of comparison with data from FIRAS on CMB spectral distortions,
where FIRAS was a sensitive device attached to the COBE satellite.

\bigskip

\noindent
Unfortunately the Bondi-Hoyle model was invented for a static object
and assumes spherically symmetric purely s-wave accretion. Studies 
of the SMBH in the giant galaxy M87 have shown since 2014 that the 
higher angular momenta strongly dominate, not surprising as the
SMBH possesses a gigantic spin angular momentum in natural units.

\bigskip

\noindent
The results from M87 suggest the upper limits
on MACHOs imposed by ROM
were too severe by some 4 or 5 orders of magnitude
and that up to 100\% of the dark matter is
permitted by arguments about CMB distortion
to be in the form of PIMBHs.

\bigskip

\noindent
{\bf Microlensing}

\bigskip

\noindent
Microlensing is the most direct experimental method and has the big advantage
that it has successfully found
examples of MACHOs. The MACHO Collaboration used a method
which had been proposed\footnote{We have read that such gravitational lensing was later found to have been calculated 
in unpublished 1912 notes by Einstein who did not publish perhaps because at that time he considered its experimental measurement
impracticable.} by Paczynski where the amplification of a distant
source by an intermediate gravitational lens is observed. The MACHO Collaboration
discovered several striking microlensing events whose light curves are
exhibited in its 2000 paper. The method certainly worked well for $M <  25 M_{\odot}$
and so should work equally well for $M > 25 M_{\odot}$ provided one can devise
a suitable algorithm and computer program to scan enough sources.

\bigskip

\noindent
The longevity of a given lensing event is proportional to the square root of the lensing mass
and numerically is given by
($\hat{t}$ is longevity)

\begin{equation}
\hat{t} \simeq 0.2 yr \left( \frac{M_{lens}}{M_{\odot}} \right)^{1/2}
\label{that}
\end{equation}
where a transit velocity $200km/s$ is assumed for the lensing object.

\bigskip

\noindent
The MACHO Collaboration [18] investigated lensing events with longevities
ranging between about two hours and one year. From Eq.(\ref{that}) this corresponds
to MACHO masses between approximately $10^{-6} M_{\odot}$ and $25 M_{\odot}$.

\bigskip

\noindent
The total number and masses of objects discovered by the MACHO Collaboration
could not account for all the dark matter known to exist in the Milky Way. At most
10\% could be explained. To our knowledge, the experiment ran out of money and
was essentially abandoned in about the year 2000.
But perhaps the MACHO Collaboration and its funding
agency were too easily discouraged.

\bigskip

\noindent
What is being suggested is that the other 90\% of the dark matter in the
Milky Way is in the form of MACHOs which are more massive than those detected
by the MACHO Collaboration, and which almost certainly could be detected by a
straightforward extension of their techniques. In particular, the expected
microlensing events have
a duration ranging up to two centuries.

\bigskip

\noindent
Microlensing experiments involve systematic scans of millions of distant star
sources because it requires accurate alignment of the star and the
intermediate lensing MACHO. Because the experiments are already highly computer
intensive, it makes us more optimistic that the higher longevity events
can be successfully analyzed. Study of an event lasting two centuries should
not necessitate that long an amount of observation time.  It does require suitably ingenious 
computer programming to track light curves and distinguish them from
other variable sources. This experiment is undoubtedly extremely
challenging, but there seems no obvious reason it is impracticable.

\bigskip

\noindent
{\bf Intermediate discussion}

\bigskip

\noindent
Axions probably do not exist for theoretical reasons
discovered in 1992. Electroweak supersymmetry probably does not exist
for the experimental reason of its non-discovery 
in Run 1 of the LHC.
The idea that dark matter experiences weak interactions (WIMPs) came historically
from the appearance of an appealing DM constituent, the neutralino,
in the theory of electroweak supersymmetry for which there is no experimental evidence.

\bigskip

\noindent
The only interaction which we know for certain to be experienced
by dark matter is gravity and the simplest assumption is that gravity
is the only force coupled to dark matter.
Why should the dark matter experience the weak interaction
when it does not experience the strong and electromagnetic interactions?

\bigskip

\noindent
All terrestrial experiments searching for dark matter by either direct
detection or production may be doomed to failure. 

\bigskip

\noindent
We began with four candidates for dark matter constituent:
(1) axions; (2) WIMPs; (3) baryonic MACHOs;
(4) PIMBHs.
We eliminated the first two by hopefully persuasive
arguments, made within the context of an overview
of particle phenomenology including a combination of old and new
results. We eliminated the third by the upper limit
on baryons imposed by robust Big Bang 
Nucleosynthesis (BBN) calculations.

\bigskip

\noindent
We assert that PIMBHs can constitute almost all  dark matter while maintaining
consistency with the BBN calculations. This is an important point because distinguished
astronomers have written an opposite assertion {\it e.g.} Begelman and Rees 
state that black holes cannot form more than 20 \% of dark matter
because the remainder is non-baryonic. 

\bigskip

\noindent
These authors are making an implicit assumption which does not apply to 
the PIMBHs which we assert comprise almost all dark matter. That assumption is that
black holes can be formed only as the result of the gravitational collapse
of baryonic stars. We are claiming, on the contrary,  that dark matter black holes can be,
and the majority must be,
formed primordially in the early universe as calculated and demonstrated in 
FKTY(2010) and independently by CKSY(2010).

\bigskip

\noindent
Our proposal is that the Milky Way contains between ten million and ten billion
massive black holes each with between a hundred and a hundred thousand times the solar mass. Assuming the halo
is a sphere of radius a hundred thousand light years the typical separation
is between one hundred and one thousand light years which is also the most
probable distance of the nearest PIMBH to the Earth. At first sight, it may be
surprising that such a huge number of PIMBHs

\noindent
-- the plums in a {\it ``PIMBH plum pudding"} --

\noindent
(c.f. Thomson 1904) could remain undetected.

\noindent
[111 years after Thomson; 31 powers of ten bigger;
not replaceable by a nuclear halo.]

\bigskip

\noindent
However, the mean
separation of the plums
is at least a hundred light years and the plum size
is smaller than the Sun.

\bigskip

\noindent
{\bf PIMBHs and PSMBHs}

\bigskip

\noindent
Focusing on the Milky Way halo where we can most easily detect the PBHs, we
already know from earlier searches, especially the MACHO Collaboration
that masses $M \leq 20M_{\odot}$ can make
up no more that 10\% of the halo dark matter.  At the high mass end,
we know from Xu-Ostriker that MACHOs with $M \geq 10^5 M_{\odot}$
endanger disk stability. For the Milky Way halo one is led
to consider intermediate mass PIMBHs in the mass range

\begin{equation}
20 M_{\odot} \leq M_{PIMBH} \leq 10^5 M_{\odot}
\label{PIMBHmass}
\end{equation}
 
 \noindent
for the DM constituents. This leads to a {\it plum pudding} model 
for the
Milky Way halo, named after Thomson's atomic model, where
for the DM halo the plums are PIMBHs with masses satisfying Eq.(\ref{PIMBHmass}) and
the pudding is rarefied gas, dust and a few luminous stars.

\bigskip

\noindent
The formation of PBHs with masses as large as Eq.(\ref{PIMBHmass}) and much larger
is known to be mathematically possible during the radiation era. An existence theorem
is provided by hybrid inflationary models. One specific prediction of hybrid
inflation is a sharply-peaked PBH mass function. If we need a specific PIMBH mass, 
we shall use a calligraphic ${\cal PIMBH}$ defined by  $M_{{\cal PIMBH}} \equiv 100 M_{\odot}$
exactly. This is merely an example and extension to the whole range of Eq.(\ref{PIMBHmass})
can also be discussed.

\bigskip

\noindent
The cosmic time $t_{PBH}$ at which a PBH is formed has been estimated to be

\begin{equation}
t_{PBH} \simeq \left( \frac{M_{PBH}}{10^5M_{\odot}} \right) ~~ seconds
\label{tPBH}
\end{equation}

\noindent
so that the PIMBHs in Eq.(\ref{PIMBHmass}) are formed in the time window
$0.0002s \leq t_{PIMBH} \leq 1.0s$
with the special case $t_{{\cal PIMBH}} \simeq 0.001 s$. In terms of red shift ($Z$),
this corresponds to
\begin{equation}
5 \times 10^{11} \geq Z_{PIMBH} \geq 5 \times 10^9
\label{ZPIMBH}
\end{equation}
with the special case $Z_{{\cal PIMBH}} \simeq 2 \times 10^{11}$. 

\bigskip

\noindent
The formation of BHs which are not primordial, which we shall
denote without an initial $P$ or ${\cal P}$, necessarily occurs {\it after} star formation which conservatively
occurs certainly only for very different redshifts satisfying 

\begin{equation}
Z_{BH} \leq 100
\label{ZBH}
\end{equation}

\noindent
The sharp difference in the red-shifts of Eq.(\ref{ZPIMBH}) and Eq.(\ref{ZBH}) will become 
important when we discuss the reasons for previous non-detection, the angular momentum of PIMBHs and BHs,
and the central issue of possible CMB distortion by X-rays.

\bigskip

\noindent
As already mentioned, by using the mathematical models, it is possible to form PBHs 
not only in the PIMBH mass range of Eq. (\ref{PIMBHmass}) but also Primordial Super Massive
Black Hole (PSMBHs) in the mass range
\begin{equation}
10^5 M_{\odot} \leq M_{PSMBH} \leq 10^{17} M_{\odot}
\label{PSMBHmass}
\end{equation}
where the upper limit derives from the formation time $t_{PSMBH}$ given by Eq. (\ref{tPBH}) staying
within the radiation-dominated era. We shall discuss the higher mass range Eq( \ref{PSMBHmass}) later in the paper.

\bigskip

\noindent
Finally for this Introduction, we recall that in a microlensing experiment, {\it e.g.} using the LMC or SMC for 
convenient sources, microlensing by halo PIMBHs, and assuming a typical transit velocity $200km.s^{-1}$, the time duration
of the microlensing light curve can be estimated to be approximately
\begin{equation}
\tau \simeq \left( \frac{M_{PIMBH}}{25 M_{\odot}} \right)^{\frac{1}{2}}  ~~ years
\label{duration}
\end{equation}
which we note is close to one year and two years, respectively, for lens masses 
$25M_{\odot}$ and $100 M_{\odot}$. For reference, the highest duration such light curve
detected by the MACHO Collaboration which published in the year 2000 
corresponded to $M_{PIMBH} \simeq 20 M_{\odot}$.

\bigskip

\noindent
Nevertheless, if longer duration microlensing light curves can be detected of two years or more,
the only known explanation will be the existence of Kerr black holes in the halo with many
solar masses.

\bigskip

\noindent
{\bf Kerr metric and period $\tau$}

\bigskip

\noindent
The PIMBHs are described by a Kerr metric which has the form in 
Boyer-Lindquist $(t, r, \theta, \phi)$ coordinates, after defining
$\alpha = \frac{J}{M}$, $\rho^2 = r^2 + \alpha^2 \cos^2 \theta$ and $\Delta = r^2-2Mr +\alpha^2$,

\begin{eqnarray}
ds^2 &=& - \left( 1 - \frac{2Mr}{\rho^2} \right) dt^2 - \left( \frac{4Mr\alpha \sin^2\theta}{\rho^2} \right) d\phi dt \nonumber \\
& &+ \left( \frac{\rho^2}{\Delta} \right) dr^2 + \rho^2 d\theta^2 \nonumber \\
& & +\left( r^2 + \alpha^2 + \frac{2Mr \alpha^2 \sin^2 \theta}{\rho^2} \right) \sin^2 \theta d\phi^2
\label{Kerr}
\end{eqnarray}

\bigskip

\noindent
In Eq.(\ref{Kerr}), there are two free parameters, $M$ and $J$. Analytic calculations
building on Eq. (\ref{Kerr}) can be difficult, usually leading to numerical techniques.

\noindent
In this talk, we shall need only order-of-magnitude estimates for the rotational period $\tau$ and, in the
next Section, for the angular momentum $J$. These will suffice to make our point about concomitant
X-ray emission. The solution is axially symmetric and the radius at the pole $\theta = \frac{\pi}{2}$
is the same as the Schwarzschild radius $R = 2M$. For other values of $\theta$ the black hole radius
is smaller than the static one and the rest of the static would-be sphere is filled out by an ergosphere whose equatorial
radius is also $R=2M$.  

\bigskip

\noindent
For the primordial black holes of interest, there is no reason to expect that the radiation
will collapse in a spherically symmetric fashion to a static Schwarzschild black hole when
the PBH formation necessarily occurs in an  environment of extreme fluctuations and 
inhomogeneities. The black holes must be described by the
Kerr metric in Eq.(\ref{Kerr}) with $\alpha$ having a value anything up to the maximal
Kerr solution which corresponds to an equatorial speed $V$ equal to the speed of light.
The range of $V$ is thus $0 \leq V \leq c$.

\bigskip

\noindent
We do not know observationally any black hole which is primordial with certainty although
many of the observed black holes, including those in the binary coalescences observed
by LIGO, could be primordial. For illustration
of black hole observations, let us consider the well-studied binary GRS1915+105
of a star and a black hole. 

\bigskip

\noindent
The black hole mass in GRS1915+105 has been established
as $M \simeq 13 M_{\odot}$ and hence its \\
Schwarzschild radius is $r_s \simeq 39$ km.
Its rotation occurs $1,150$ times per second which translates
to an equatorial speed $V\simeq 0.94c$, remarkably close to maximal. We mention
this example to show that such high $V$ Kerr black holes are known to
exist and although we cannot derive the value of $V$ arising from PBH formation
it is to be expected that all values $V$ up to the maximum can occur. For
our present qualitative purposes, to be conservative, we employ $V=0.1c$. 

\bigskip

\noindent
To proceed with our estimate we shall therefore
take the equatorial velocity of the ergosphere to have magnitude $V = 0.1c$
and use Newtonian mechanics to estimate the rotation period $\tau$ as simply

\begin{equation}
\tau = \left( \frac{2\pi R}{V} \right)
\label{period}
\end{equation}

\noindent
For the Sun, we have $2M_{\odot} \simeq 3 ~ km$ so that for a black hole of mass 
$M = \eta M_{\odot}$ and therefore radius $R \simeq 3 \eta ~ km$ Eq.(\ref{period}) is, for
$V=0.1c = 3 \times 10^4 km.s^{-1}$,

\begin{equation}
\tau = \left( 2 \times 10^{-4} \pi \eta \right) ~~ seconds
\label{tau}
\end{equation}

\bigskip

\noindent
Some values of $\tau$, estimated by this method, are shown in the third column of 
the Table and angular momentum {\cal J} (discussed later) is in the last column.

\vspace{1.0in}

\begin{table}
\begin{center}
\begin{tabular}{|c|c|c|c|}
\hline
Astrophysical & Mass  & Period $\tau$   & Angular Momentum ${\cal J}$ \\
object       &  solar masses  & seconds  & $kg.km^2.s^{-1}$    \\
\hline
\hline
Earth & $M_{\oplus} = 6 \times 10^{24} kg$          &  24 hours                       &   $1.1 \times 10^{27}$        \\
\hline
Sun &   $M_{\odot} = 2 \times 10^{30} kg$      &    25 days           &  $1.1 \times 10^{36}$       \\
\hline
\hline
PIMBH & $20 M_{\odot}$ & $0.013 s $ & $3.0 \times 10^{37}$  \\
\hline
${\cal PIMBH}$ &  {\cal $100 M_{\odot}$} & {\cal $0.063 s$} &  $7.2 \times 10^{38}$  \\
\hline
PIMBH & $1000 M_{\odot}$  & $0.63s$ &  $7.2 \times 10^{40}$ \\
\hline
PIMBH & $10^4 M_{\odot}$  & $6.3s$ &  $7.2 \times 10^{42}$ \\
\hline
PIMBH & $10^5 M_{\odot}$ & $63s$ & $7.2 \times 10^{44}$ \\
\hline
\hline
PSMBH (M87) & $6 \times 10^9 M_{\odot}$ & $3.8 \times 10^6 s$ & $2.6 \times 10^{54}$  \\
\hline
\hline
\end{tabular}
\end{center}
\label{tauJ}
\end{table}

\bigskip

\noindent
{\bf Angular momentum ${\cal J}$}

\bigskip

\noindent
Let us define the dimensionless angular momentun ${\cal J} \equiv J / kg.km^2.sec^{-1}$. We
are interested in order of magnitude estimates of ${\cal J}$ for the PIMBHs and PSMBHs. The value of ${\cal J}$ for astrophysical objects is necessarily
a large number so to set the scene we shall estimate ${\cal J}$ for the Earth ${\cal J}_{\oplus}$
and for the Sun ${\cal J}_{\odot}$.

\bigskip

\noindent
The parameters for the Earth are radius $R_{\oplus} \simeq 6300 km$, period $\tau_{\oplus} \simeq 86400 s$, mass
$M_{\oplus} \simeq 6 \times 10^{24} kg$,
hence angular velocity $\omega_{\oplus} = 2\pi /\tau_{\oplus}$ and moment of inertia $I_{\oplus} = \frac{2}{5} M_{\oplus}R_{\oplus}^2$
so an estimate is ${\cal J}_{\oplus} \sim I_{\oplus} \omega _{\oplus} \simeq 1.1 \times 10^{27}$.  For the Sun the similar calculation
using $R_{\odot} \simeq 700,000 km$, $\tau_{\odot} \simeq 25 days$, $M_{\odot} \simeq 2 \times 10^{30} kg$ gives ${\cal J}_{\odot} \simeq 1.1 \times 10^{36}$.

\bigskip

\noindent
For the black holes, the value of ${\cal J}$ is proportional to $\eta^2$ where $\eta = (M / M_{\odot})$. A similar estimate
to that for the Earth and Sun gives ${\cal J} \simeq 7.2 \times 10^{34} \eta^2$, which provides the remaining entries in the Table.

\bigskip

\noindent
{\bf CMB distortion revisited}

\bigskip

\noindent
Because of rotational invariance, angular momentum is conserved. The ${\cal J}$
of a compact astrophysical object will not change dramatically unless there is
an extremely unlikely event like a major collision. For example, the Earth and the
Sun in the first two rows of our Table were formed $4.6$ billion years ago. Their 
respective  angular momenta ${\cal J}_{\oplus}$ and ${\cal J}_{\odot}$ have
remained essentially constant all of that time. According to Eq.(\ref{tPBH}),
the PIMBHs listed in the next five rows of our Table were all formed at time
$t \leq 1s$ and their angular momenta have therefore remained roughly constant for
the last $13.8$ billion years since then.

\bigskip

\noindent
In detecting the dark matter, let us focus on the special case ${\cal PIMBH}$
with $M=100M_{\odot}$. The ${\cal PIMBH}$ was formed, accordng to
Eq.(\ref{tPBH}), at time $t=10^{-3} s$ and rotates with period $t\simeq 63ms$,
thus rotating $\sim 16$ times per second and with an absolute angular momentum
$\sim 6 \times 10^{11}$ times that of the Earth and $\sim 600$ times that of the Sun.
There is no known reason that ${\cal J}_{{\cal PIMBH}}$ would change
significantly after its formation.

\bigskip

\noindent
These remarks about angular momentum
are salient to resolving the contradiction between
the PIMBH dark matter proposal and the limits on halo MACHOs
derived earlier by Ricotti, Ostriker and Mack (ROM) on 
the basis of X-ray emission and CMB distortion.

\bigskip

\noindent
The PIMBH proposal was made that the Milky Way dark halo
is a plum pudding with, as ``plums", PIMBHs in the mass range of Eq.(\ref{PIMBHmass})
making up $100\%$ of the dark matter. On the other hand,
in Figure 9 of ROM, there is displayed an upper limit of
less than $0.01\%$ of the dark matter for this mass range
of MACHO. Thus, it would seem that at least one must be incorrect? 
The conclusion of the present talk is
that ROM is correct for stellar-collapse black holes
but is not applicable to a model  which
employs primordial black holes. 

\bigskip

\noindent
This ROM upper limit arises from the lack of any observed departure
of the CMB spectrum from the predicted black-body curve or of any
CMB anisotropy. ROM calculated the accretion of matter on to the
MACHOs, the emission of X-rays by the accreted matter and then the
downgrading of these X-rays to microwaves by cosmic expansion
and more importantly by Compton scattering from electrons.

\bigskip

\noindent
A crucial assumption made by ROM is that the accretion on to
the MACHO can be modeled as if the MACHO has zero angular momentum
$ J = 0$. The justification for this assumption is based on 
earlier work by Loeb [19] who studied the collapse of gas clouds
at redshifts $200 \leq Z \leq 1400$. Such collapse can form compact
objects, eventually black holes, but during the collapse angular
momentum is damped out from the electrons by Compton scattering
with the CMB.

\bigskip

\noindent
From Loeb's discussion, the resultant black holes will have $J=0$
and this appears to underly why ROM used the Bondi-Hoyle
model which presumes spherical symmetry for accretion.
This is justified for stellar-collapse black holes by the arguments
of Loeb and therefore the upper bounds derived by ROM
are applicable.

\bigskip

\noindent
There is evidence that the Bondi-Hoyle model of
accretion is not, by contrast, applicable to spinning PSMBHs, in particular the
one at the centre of the large galaxy M87. In recent analyses
Bondi-Hoyle was used to calculate the number
of X-rays expected from the accreted material near M87.
In the case of M87 the X-rays
are experimentally measured. The conclusion is striking: that the
measured X-rays are less by several orders of magnitude than
predicted by Bondi-Hoyle theory.

\bigskip

\noindent
This supports the idea that the SMBHs such as that in M87
are primordial, so we list PSMBH(M87) in the final row of 
our Table. The ROM
constraints apply to black holes which originate from gravity 
collapse of baryonic stars. Collecting this fact, together with the
ROM limit of $\leq 10^{-4}$ of the dark matter for MACHOs, implies that
$99.99\%$ of the dark matter black holes are primordial, formed
during the radiation era.

\bigskip

\noindent
{\bf Final discussion}

\bigskip

\noindent
The dark matter and its explanation is a pressing problem which impacts on
both high-energy physics and on cosmology. It is indisputable that over
80\% of the Milky Way's mass lies in a dark approximately spherical halo
surrounding the luminous more planar spiral. The results in the present
Letter strongly support the model involving billions of PIMBHs.

\bigskip

\noindent
The plum pudding model for the dark halo proposed in 
arose from a confluence of theoretical threads including study of the entropy
of the universe and the knowledge of how to form PBHs with 
many solar masses as in Eqs. (\ref{PIMBHmass}) and (\ref{PSMBHmass}). Nevertheless it was
the weakening of the argument for WIMPs which was most
decisive,

\bigskip

\noindent
The strongest objection to the PIMBHs has been based on
the X-rays and the CMB distortion 
as calculated by ROM. In the present talk we have
attempted to lay this criticism to rest by noting that ROM assumed $J=0$
and that the putative PIMBHs have not only many times the Solar mass
but also many times the Solar angular momentum. This appears to us to render
the ROM constraints inapplicable to the PIMBHs. On the other hand, they
do apply to stellar-collapse black holes which implies that almost none
($\leq 0.01\%$) of the dark matter black holes are of that type.
To decide whether dark matter really is PIMBHs will require their
detection by a dedicated microlensing experiment.

\bigskip

\noindent
Examples
of PSMBHs may already have been observed in galactic cores and
quasars. Other PSMBHs can play the role of dark matter in clusters
and may well be detectable by other future lensing experiments. 
There is
also the upper mass range contained in Eq.(\ref{PSMBHmass}). 
Although masses of PSMBHs up to a few times
$10^{10} M_{\odot}$ may have already been observed in quasars, there
are what could be called Primordial Ultra Massive Black Holes (PUMBHs)
with masses between $10^{11}$ and $10^{17}$ solar masses which
might exist within the visible universe.

\bigskip

\noindent
PUMBHs remain speculative but what can in the near future be examined 
experimentally is the existence of PIMBHs in the halo.
A positive result would solve the 83-year-old problem of the dark matter and explain $\sim 26.7\%$
of the total stress-energy tensor of the visible universe. It would presumably put a stop to searches
for WIMPs because the scientific community would accept that
WIMPs, like low-energy supersymmetry, do not exist. Searches for axions would
perhaps continue but purely within the particle physics domain with no notion that axions, 
if they exist, can form more than a very tiny fraction of dark matter.

\bigskip

\noindent
The identification of the dark matter constituents as PIMBHs can revolutionize astronomy and cosmology. To give just one example, 
the formation
of stars which takes place at redshifts $Z \leq 8$ becomes as if only a minor ``afterthought"
with regard to all the earlier large scale structure  formation which would take place in a Universe
containing {\it only} dark matter in the form of PIMBHs. In this sense, the result of this experiment can diminish the cosmological significance of normal matter.

\bigskip

\noindent
Of the detection methods discussed, extended microlensing observations
seem the most promising and an experiment to detect
higher longevity microlensing events is being actively pursued.
The wide-field telescope must be in the Southern Hemisphere
to use the Magellanic Clouds (LMC and SMC) for sources.

\bigskip

\noindent
The most appropriate active telescope has been identified as the
Blanco 4m at Cerro Tololo in Chile.                                                                                                                                                                                                                                                                                                                                This telescope was named after the late Victor Blanco
the Puerto Rican astronomer who was the CTIO Director.
A bigger telescope which can confirm the high-duration
light curves is the Large Synoptic Survey Telescope (LSST)
under construction, also in Chile, expected to take first light
in 2022.

\vspace{0.5in}

\newpage

\noindent
{\bf REFERENCES}

\bigskip

[1] D.W. Sciama, {\it Modern Cosmology and the Dark Matter}. Cambridge University Press (2008).

[2] R.H. Sanders, {\it The Dark Matter Problem, A Historical Perspective}. Cambridge Uni-versity Press (2014).

[3] G. Bertone (Editor), {\it Particle Dark Matter, Observations, Models and Searches}. Cambridge University Press (2013).

[4] K. Freese, {\it The Cosmic Cocktail, Three Parts Dark Matter}. Princeton University Press (2014).

[5] P.H. Frampton, {\it Searching for Dark Matter Constituents with Many Solar Masses}.
Mod. Phys. Lett. {\bf A31,} 1650093 (2016). {\tt arXiv:1510.00400}.\\
G. Chapline and P.H. Frampton,\\
{\it A New Direction for Dark Matter Research: Intermediate Mass Compact Halo Objects}.\\
JCAP {\bf  11,} 042 (2016). {\tt arXiv:1608.04297[gr-qc]}.

[6] Y.B. Zel'dovich and I.D. Novikov, \\
Astron. Zh. {\bf 43,}758 (1966).

[7] B.J. Carr and S.W. Hawking,\\
{\it Black Holes in the Early Universe}.\\
Mon. Not. Roy. Astron. Soc. {\bf 168,} 399 (1974).

[8] G.F. Chapline,\\
{\it Cosmological Effects of Primordial Black Holes}.\\
Nature {\bf 253,} 251 (1975).

[9] P.H. Frampton, M. Kawasaki, F. Takahashi and T.T. Yanagida, \\
{\it Primordial Black Holes as All Dark Matter}.\\
JCAP 1004 023 (2010). {\tt arXiv:1001.3208[hep-ph]}.

[10] S. Clesse and J. Garcia-Bellido,\\
{\it Massive Primordial Black Holes from Hybrid Inflation
as Dark Matter and the Seeds of Galaxies}.\\
Phys. Rev. {\bf D92,} 023524 (2015). {\tt arXiv:1501.07565[astro.CO]}.

[11] B.P. Abbott, {\it et al.} (LIGO Scientific and Virgo Collaborations).\\
 {\it Observation of Gravitational Waves from a Binary Black Hole Merger}.\\
 Phys. Rev. Lett. {\bf 116,} 061102 (2016).
 {\tt arXiv:1602.03837[gr-qc]}.

[12] S. Bird, I. Cholis, J.B. Munoz, Y. Ali-Haimoud, M. Kamionkowski,
E.D. Kovetz, A. Raccanelli and A.G. Riess,\\
{\it Did LIGO Detect Dark Matter?}\\
Phys. Rev. Lett. {\bf 116,} 201301 (2016). {\tt arXiv:1603.00464[astro-ph.CO]}.

[13] S. Coleman,\\
{\it Fate of the False Vacuum: Semiclassical Theory}.\\
Phys. Rev. {\bf D15,} 2929 (1977).

[14] P.H. Frampton,\\
{\it Vacuum Instability and Higgs Scalar Mass}.\\
Phys. Rev. Lett. {\bf  37,} 1378 (1976).

[15] J. Yoo, J. Chaname and A. Gould, \\
{\it The End of the MACHO Era: Limits on Halo Dark Matter from Stellar Halo Wide Binaries}. \\ Astrophys. J. {\bf 601,} 311 (2004). {\tt arXiv:astro-ph/0307437}.

[16] D.P. Quinn, M.I. Wilkinson, M.J. Irwin, J. Marshall, A. Koch \\
and V. Belokurov, \\
{\it On the Reported Death of the MACHO Era}. \\ Mon. Not. Roy. Astron. Soc. {\bf 396,} 11 (2009). 
{\tt arXiv:0903.1644[astro-ph.GA]}.

[17] M. Ricotti, J.P. Ostriker and K.J. Mack, \\
{\it Effect of Primordial Black Holes on the Cosmic Microwave Background and Cosmological Parameter Estimates}. \\
Astrophys. J. {\bf 680,} 829 (2008). {\tt arXiv:1709.0524[astro-ph]}.

[18] C. Alcock, {\it et al.}, \\
{\it The MACHO Project: Microlensing Results from 5.7 Years of 
 LMC Observations.} 
 Astrophys. J. {\bf 542,} 281 (2000). 
 {\tt arXiv:astro-ph/0001272 }. 
 
 [19] A. Loeb, {\it Cosmological Formation of Quasar Black Holes.} \\
 Astrophys. J. {\bf 403,} 542 (1993).

\end{document}